\documentclass[%
reprint, 
amsmath,amssymb,
aps,
nofootinbib
prb,
superscriptaddress
]{revtex4-1}

\usepackage{scrextend}
\usepackage{graphicx}
\usepackage{dcolumn}
\usepackage{bm}
\usepackage{tabularx}
\usepackage[utf8]{inputenc}
\usepackage[english]{babel}
\usepackage{color}

\usepackage{physics}

\usepackage{amsmath}
\usepackage{soul}
\DeclareMathAlphabet{\pazocal}{OMS}{zplm}{m}{n}

\usepackage{hyperref}

\begin{document}
	
\preprint{APS/123-QED}
\title{Lead-free Magnetic Double Perovskites for Photovoltaic and Photocatalysis Applications}
\author{Muskan Nabi}
\affiliation{Department of Physics, Indian Institute of Technology, Bombay, Powai, Mumbai 400076, India}
\author{Sanika S. Padelkar}
\affiliation{Department of Physics, Indian Institute of Technology, Bombay, Powai, Mumbai 400076, India}
\affiliation{School of Chemistry, Monash University, Victoria 3800, Australia}
\affiliation{IITB-Monash Research Academy, IIT Bombay, Mumbai 400076, India}
\author{Jacek J. Jasieniak}
\affiliation{Department of Materials Science \& Engineering, Monash University, Victoria 3800, Australia}
\author{Alexandr N. Simonov}
\affiliation{School of Chemistry, Monash University, Victoria 3800, Australia}
\author{Aftab Alam}
\email{aftab@iitb.ac.in}
\affiliation{Department of Physics, Indian Institute of Technology, Bombay, Powai, Mumbai 400076, India}
\affiliation{IITB-Monash Research Academy, IIT Bombay, Mumbai 400076, India}

\begin{abstract}
The magnetic spin degrees of freedom in magnetic materials serve as additional capability to tune materials properties, thereby invoking magneto-optical response. Herein, we report the magneto-optoelectronic properties of a family of lead-free magnetic double perovskites Cs$_{2}$AgTX$_{6}$ (T = Sc, Ti, V, Cr, Mn, Fe, Co, Ni, Cu; X=Cl, Br, I). This turns out to provide an extremely fertile series, giving rise to potential candidate materials for photovoltaic(PV) applications. In conjunction with high absorption coefficient and high simulated power conversion efficiency for PV applications, few compounds in this series exhibit novel magnetic character useful for spintronic applications. The interaction between magnetism and light can have far-reaching results on the photovoltaic properties as a consequence of the shift in the defect energy levels due to Zeeman effect. This subsequently affects the recombination rate of minority carriers, and hence the photoconversion efficiency. Moreover, the distinct ferromagnetic and anti-ferromagnetic ordering driven by hybridization and super-exchange mechanism can play a significant role to break the time-reversal and/or inversion symmetry. Such a coalescence of magnetism and efficient optoelectronic response has the potential to trigger magnetic/spin anomalous photovoltaic (non-linear Optical) effect in this Cs$_{2}$AgTX$_{6}$ family. These insights can thus channelize the advancement of lead-free double perovskites in magnetic/spin anomalous photovoltaic field as well.
\end{abstract}


\maketitle
\section{Introduction} 
Organic-inorganic halide double perovskites have emerged as a promising class of materials in various fields such as ferroelectrics,\cite{key_1} spintronics,\cite{key_2} photovoltaics,\cite{key_3,key_4} and optoelectronic devices such as light emitting diodes (LED’s),\cite{key_5}  sensors,\cite{key_6}, X-ray detectors\cite{key_7} and photo-detectors {\cite{key_8}}. Lead-free halide double perovskites (DPs) with general formula  A$_{2}$BB$'$X$_{6}$, formed by a combination of one monovalent and one trivalent ion, have emerged ubiquitously as a stable and green alternative to toxic lead-based halide perovskites. The adept optoelectronic properties of these materials can be associated with the compositional flexibility, dielectric properties and exciton binding energies ranging over several orders of magnitude \cite{key_9,key_10}. Amongst the DPs family, materials with A=Cs$^{+1}$, B= Ag$^{+1}$ or Cu$^{+1}$ and B$'$=Bi$^{3+}$,Sb$^{3+}$ or In$^{3+}$  have been proposed to be environmentally friendly alternatives to lead-based perovskites \cite{key_11}. Some of the experimentally studied materials, e.g.,  Cs$_{2}$AgBiX$_{6}$ (X=Cl, Br, I)\cite{key_12,key_13},  Cs$_{2}$AgSbX$_{6}$\cite{key_14,key_15}, Cs$_{2}$AgInX$_{6}$ \cite{key_16,key_17} and Cs$_{2}$InB$^{3+}$X$_{6}$ (B$^{3+}$ =Sb, Bi) \cite{key_18,key_19} have received substantial interest because of their promising properties. Cs$_{2}$AgBiBr$_{6}$ \cite{key_20} is one of the most frequently investigated materials in this class having higher thermodynamical stability although with an indirect band gap of about 2 eV and power conversion efficiency of about ca 3\%. However, recent experimental and theoretical evidences have shown that the Ag-Bi variants exhibit intrinsic and strong electronic confinement, which is manifested in very large exciton binding energies (hundreds of meV), strong carrier localization and reduced free-carrier mobility\cite{key_21,key_22}. The exciton binding energies in halide DPs are influenced by the electronic structure of the alternating B and B$'$ site cations. Hence, via chemical substitution at the B and B$'$  sites, existing set of halide DPs can be optimized towards better performance. The seminal contributions to modulate the existing class of materials have been underpinned by several viable strategies like alloy/doping mediated band-gap engineering, optimizing synthesis processes with a view to tackle critical challenges \cite{key_23}.

However, an aspect that remains underexplored but presents a lot of promise is the magnetic spin degrees of freedom available in magnetic perovskites to tune the photovoltaic (PV) properties, which in all likelihood can lead to staggering spin-related properties. In this regard, few halide perovskites, viz.,Cs$_{2}$AgT$^{3+}$Cl$_{6}$ (T=Fe, Cr) \cite{key_24, key_25}, Cs$_{2}$NaT$^{3+}$Cl$_{6}$ (T=Fe, V, Mn, Ni) \cite{key_26, key_27} , Cs$_{2}$KT$^{3+}$Cl$_{6}$ (T=Mn, Co, Ni) \cite{key_28}, Cs$_{2}$GeT$^{3+}$X$_{6}$ (M = Ti, V, Cr, Mn, Fe, Co, Ni, or Cu)\cite{key_29}  and various oxide perovskites \cite{key_30, key_31} have been reported to show interesting magnetic properties. Amongst these compounds, cubic Cs$_{2}$AgFeCl$_{6}$ \cite{key_24} in particular has been experimentally reported to exhibit promising optoelectronic characteristics and PV performance. Correspondingly, hexagonal Cs$_{2}$AgCrCl$_{6}$ \cite{key_25} was synthesized in the paramagnetic phase exhibiting appreciable optoelectronic properties. Yet, the interconnection between the magnetic degrees of freedom and the optical properties are still due to be explored. The interplay of two spin degrees of freedom in these magnetic systems gives a wider plat-form for modulating the absorption range with an attempt to harvest the entire solar radiation spectrum.

Oxide perovskite Bi$_{2}$FeCrO$_{6}$ {\cite{key_32} is another state-of-the-art magnetic material which has been experimentally reported with remarkable efficiency of ca 8\%. The interaction between the magnetic field and light is called the magneto-optical effect, which includes - Zeeman effect, Faraday effect and magneto-optical kerr effect. The exceptional performance of Bi$_{2}$FeCrO$_{6}$ is attributed to two prime magneto-optical effects, viz., magnetoelectric coupling and Zeeman effect. This system has gained attraction on account of its coexistent ferroelectric properties with inbuilt polarization - thus leading to magnetoelectric coupling. Whereas, the Zeeman effect causes the energy levels to split when placed in an applied magnetic field. As a consequence of which, the energy levels depart from the band gap center, reducing the recombination rate of minority carriers, and thus prolonging minority-carriers lifetime. This capability of magnetism in tuning the optoelectronic properties is widely explored in oxide perovskites and not much has been reported for halide perovskites. The latter are good photovoltaic materials with promising photovoltaic power conversion efficiencies \cite{key_33}. So, combining the magnetic effect with halide perovskites can provide new opportunities for highly efficient devices. Herein, we primarily focus to propose a family of halide Dps where an amalagam of magnetic and optoelectronic properties will be discussed towards magneto-photovltaic applications. With that perspective, we studied a set of 27 compounds Cs$_{2}$AgT$^{3+}$X$_{6}$ (T=Sc,Ti,V,Cr,Mn,Fe,Co,Ni,Cu; X=Cl,Br,I) using the \emph{ab-initio} density functional  theory (DFT) simulation to explore the interplay between different magnetic ordering and optoelectronic properties. For example, a magnetic semiconductor with unequal band gap in the two spin channels can capture two different ranges of the solar spectrum and hence has the capability to provide higher quantum yield. A detailed structural and chemical phase stability calculation confirms 14 out of 27 compounds to be stable in a single phase, each with different magnetic ordering. Few others also show robust stability but with the possibility for the formation of a secondary phase. Due to the presence of 3d transition elements (T$^{3+}$), this family of compounds gives a fertile ground to realize several interesting properties such as half metallic ferromagnets, antiferromagnetic semiconductors, ferromagnetic and nonmagnetic semiconductors. Based on these versatile properties, these compounds are classified for different renewable energy applications. Such a detailed study of synthesizability, electronic/magnetic structure and optoelectronic properties paves a guiding path to experimentalists for future exploration these novel magnetic perovskites.

\section{Computational Details}
All calculations are performed using the DFT as implemented in the Vienna ab-initio simulation package (VASP) \cite{key_34,key_35}. For spin-polarized calculations, electronic exchange correlation functional Perdew-Burke-Ernzerhof (PBE) \cite{key_36} were used within generalized gradient approximation (GGA) \cite{key_37} along with projected augmented wave (PAW) pseudopotentials \cite{key_38,key_39}. For the wavefunction expansion, a plane-wave energy cut-off of about 450 eV was used for all calculations. The Brillouin zone integration was done within the tetrahedron method using 8$\times$8$\times$8 k-point grid for structural optimization. While for self-consistent field (SCF) calculations, {$\Gamma$-centered k-mesh} of size 12$\times$12$\times$12 is used. Keeping in mind the possibility of both cubic and hexagonal phases (as experimentally reported for two different systems belonging to this class) along with different magnetic ordering, structural optimization for all the compounds was carried out in both these phases considering non-magnetic (NM), ferromagnetic (FM) and anti-ferromagnetic (AFM) ordering. The force (energy) was converged up to 10$^{-3}$ eV\AA$^{-1}$ (10$^{-6}$ eV).

Due to the strongly correlated nature of the 3d-transition elements, an “on-site” Hubbard potential (U) \cite{key_40} was applied to capture the intra-atomic interactions between these strongly correlated electrons. U value was calculated using linear response ansatz of Cococcioni et al.\cite{key_41}. The calculation procedure and the simulated U-values for different systems are presented in Sec. S1 (see Figure S1 and Table S1) of the Supplementary Information (SI) \cite{key_42}. To estimate the theoretical photoconversion efficiency we herein reported the spectroscopic maximum limited efficiency (SLME) of the semiconducting systems \cite{key_43}.

\section{Results and Discussion}
\subsection{Structural and Chemical phase stability}

The structural stability of the halide DPs \cite{key_44} is dictated via a geometrical tolerance factor ($t$) defined as
\begin{equation}
  t=\frac{r_{A} + r_{X}}{\sqrt{2}(r_{avg}+ r_{X})}
\end{equation}
where $r_{A}$, $r_{X}$, and $r_{avg}$  are the Shannon ionic radii of A cation, X anions, and average ionic radii of B and B$'$ cations respectively. Table S2 (see supplementary) \cite{key_42} shows the tolerance factor of all the compounds which are found to lie in the range \emph{t}=0.91-0.98 predicting these compounds to stabilize in cubic structure. Interestingly, Cs$_{2}$AgCrCl$_{6}$ \cite{key_25}, was experimentally reported to crystallize in the hexagonal phase, which is normally associated with values of $t$ greater than one. Certainly, there exist other instances where the ideal cubic structure is not observed experimentally, departing from these purely geometrical tolerance factor guidelines \cite{key_45,key_46}. There are also other factors such as the B cation off-centering, and the presence and stereo activity of a lone pair of electrons which are also suggested to have strong impact on the geometrical stability of perovskites, in general \cite{key_47,key_48,key_49}. However, for transition metal based perovskites, the non-bonding (lone) pairs of electrons in transition metals do not influence molecular geometry and are said to be stereo chemically inactive. And hence we believe that the presence of lone pairs in transition metals will not affect the molecular geometry of the BX$_6$ octahedra in A$_2$BB$^{\prime}$X$_6$ DPs. To address the above mentioned ambiguities, we investigated the chemical/thermodynamics stability in both cubic as well as hexagonal phases for the entire series}.

\begin{figure}[]
	\centering
	\includegraphics[width=1\linewidth]{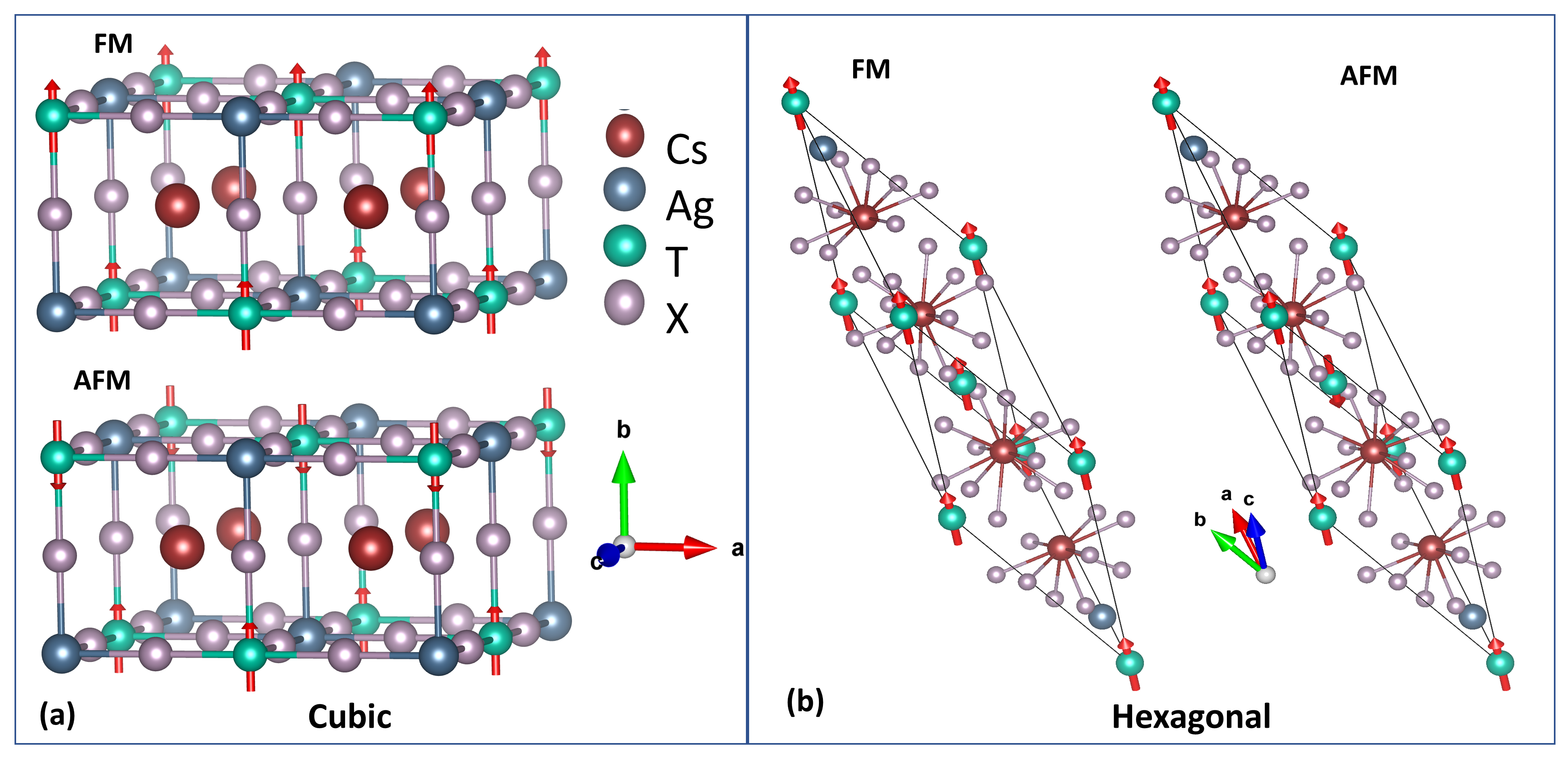}
	\caption{ Crystal structure of Cs$_{2}$AgTX$_{6}$ (T=transition elements) in (a)  cubic and (b) hexagonal phases. FM and AFM indicates ferromagnetic and antiferromagnetic ordering.}
	\label{figure-1}
\end{figure}

\begin{figure*}[]
	\centering
	\includegraphics[width=0.9\textwidth]{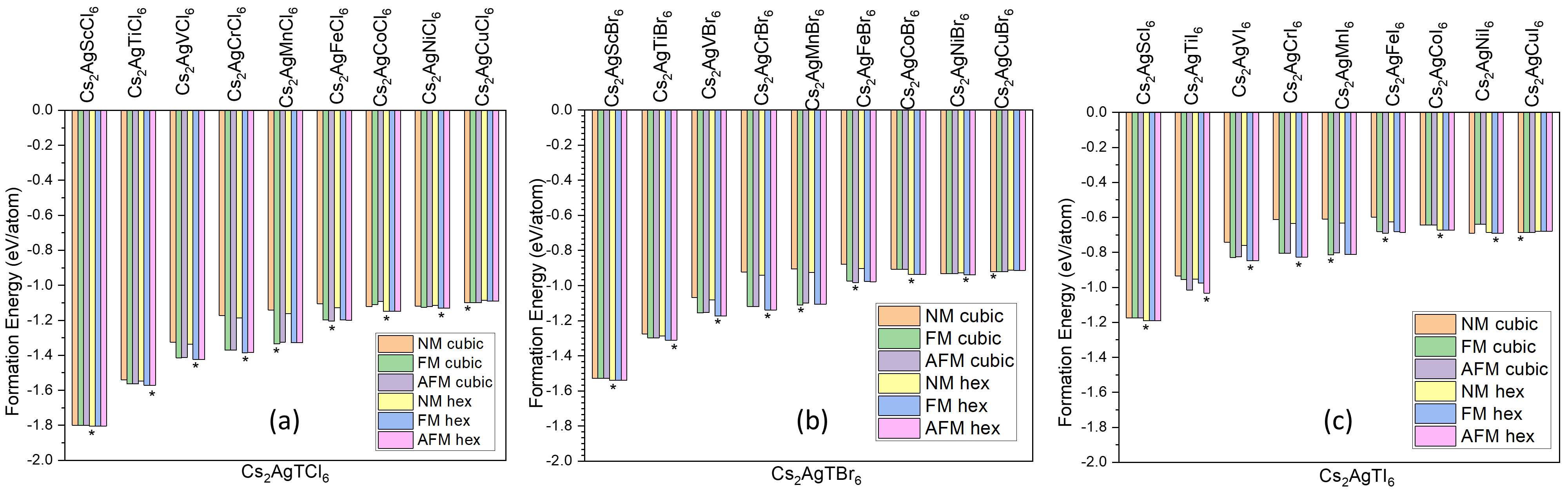}
	\caption{Formation energies ($\Delta$E$_F$) of Cs$_{2}$AgTX$_{6}$ (T=Sc,Ti,V,Cr,Mn,Fe,Co,Ni,Cu;
		X=Cl,Br,I) in two different structures (cubic and hexagonal) and three different magnetic phases (NM, FM and AFM). The asterisk (*)  indicates the energetically most stable phase for each compound.}
	\label{figure-2}
\end{figure*}

For the cubic structure of Cs$_{2}$AgTX$_{6}$, 40 atoms conventional unit cell (space group Fm$\overline{3}$m (\#225)) was considered, in which Cs is enclosed by a cage of 12 X-atoms while Ag and T adapt a corner sharing Ag$X_{6}$ and TX$_{6}$ octahedra, a s shown in Fig. \ref{figure-1}(a). The Wyckoff positions of Cs, Ag, T and X were 8c(0.25,0.25,0.25), 4b(0.5,0.5,0.5), 4a(0,0,0) and 24e(0.24,0,0) respectively. For the hexagonal phase, we used a 20 atoms primitive unit cell (space group R$\overline{3}$m (166)) adapting a Ba$_{2}$NiTeO$_{6}$ \cite{key_50} type structure, see Fig. \ref{figure-1}(b). The Wyckoff positions in this case were 6c(0,0,0.21), 6c(0,0,0.37), 3a(0,0,0) and 8h(0.48,$\overline{x}$,0.24) for Cs, Ag, T and X respectively. 
To assess the thermodynamic stability, we calculated the formation energy ($\Delta$E$_{F}$) of all the compounds in both cubic as well as hexagonal phases considering three different magnetic ordering (NM, FM and AFM), as depicted in Fig. \ref{figure-2}. The asterisk (*)  indicates the energetically most stable phase for a given compound.

The negative $\Delta$E$_{F}$ indicates the thermodynamic possibility of the formation of these compounds. Herein, a comparison of the data for materials with different halides imply that chloride-based compounds are more robust than those based on bromide and iodide. This trend validates the available experimental findings, according to which bromides and chlorides-based halide DPs are easier to synthesize than the iodide counterparts. Unsurprisingly, the experimental reports on the latter are quite rare \cite{key_51}. The actual values of formation energies of all the examined compounds in their respective stable structural/magnetic phase are provided in Table S3 of SI {\cite{key_42}.

Experimentally, {Cs$_{2}$AgFeCl$_{6}$ \cite{key_24,key_52} and Cs$_{2}$AgCrCl$_{6}$ \cite{key_25} are reported to crystallize in cubic (with anti-ferromagnetic nature at low temperature) and hexagonal structures  respectively}. Due to the involvement of 3d transition elements (Fe, Cr), these compounds are expected to show rich magnetic phase diagrams including the possibility of distinct magnetic ordering  such as FM, AFM etc. in the lower temperature range. This aspect of the aforementioned systems is overlooked in the literature but it can be extremely important to dictate their overall magneto-optical properties. As such, we have studied all the compounds in different magnetic configurations, namely NM, FM and AFM.

Remarkably, the magnetic as well as structural phases of these compounds are not significantly affected by the chemical nature of the halide anions indicating that perhaps the transition element T$^{3+}$  plays a pivotal role in determining the crystal/magnetic structure of these systems. However, the tolerance factor opens up uncertainty in the experimentally stable structural phase, as observed in case Cs$_{2}$AgCrCl$_{6}$ \cite{key_25}. Apparently, the simulated thermodynamic chemical phase diagram gives more accurate information about the degree of synthesizability of such compounds (i.e. whether in cubic/hexagonal and/or NM/FM/AFM phase). The chemical phase diagrams of the experimentally synthesized systems Cs$_{2}$AgFeCl$_{6}$ \cite{key_24}  and Cs$_{2}$AgCrCl$_{6}$\cite{key_25} are shown in Fig. \ref{figure-3}. They display a narrow stable region (shown by shaded grey area) in comparison with its competing secondary phases. For each of these two compounds, 10 secondary phases are simulated to draw their phase diagrams (Table S4) {\cite{key_42}. The narrow stable region indicates that it may be an arduous task to synthesize these systems in a single phase. Analyzing the energetics of our calculations, the most competing secondary phases for the two systems are  Cs$_{3}$Fe$_{2}$Cl$_{9}$ and Cs$_{3}$Cr$_{2}$Cl$_{9}$ which depending on the synthesis condition may restrain the respective target phases Cs$_{2}$AgFeCl$_{6}$ and Cs$_{2}$AgCrCl$_{6}$ to remain stable in a single phase. There is a possibility of the co-existence of binary phases in the synthesized compounds with a slight dominance of Cs$_{3}$Fe$_{2}$Cl$_{9}$ over Cs$_{2}$AgFeCl$_{6}$. Recently, the phase segregation and existence of secondary phases along with the target phase have also been observed in  Cs-Pb-Br thin films \cite{key_53}}. For example, Caicedo-Dávila et al. \cite{key_54} reported the coexistence of CsPbBr$_{5}$ and CsPbBr$_{3}$ during the synthesis of CsPbBr$_{5}$ based on the competing phase diagram analysis which was further validated from the DFT calculations. Also, Yu et al, \cite{key_55} confirmed that the coexistence of CsPb$_{2}$Br$_{5}$ and CsPbBr$_{3}$ is inevitable during the synthesis of CsPb$_{2}$Br$_{5}$. 
 
However, the presence of binary phases can be negated, by optimizing the synthesis process and varying the environmental parameters like pressure.There are reports where pressure is used as an important parameter to study the structural-property relationships \cite{key_56}. To crosscheck the pressure effect, we performed DFT calculations for two compounds Cs$_{2}$AgFeCl$_{6}$ and Cs$_{2}$AgCrCl$_{6}$, by systematically varying the lattice constant which in turn results in change in unit cell volume and hence the pressure. In the case of  Cs$_{2}$AgFeCl$_{6}$ and Cs$_{2}$AgCrCl$_{6}$, we have studied the chemical stability over a range of pressure from O Kb to 44 Kb and 0 Kb to 18.82 Kb respectively. The chemical phase stability diagram of these compounds in the above pressure range are presented in SI (see Fig. S3)\cite{key_42}. it can be observed that with increase in pressure the stability region of target phase decreases. Although the synthesizability of Cs$_{2}$AgFeCl$_{6}$ and Cs$_{2}$AgCrCl$_{6}$ decreases with increase in pressure, it is important to note that both Cs$_{2}$AgFeCl$_{6}$ and Cs$_{2}$AgCrCl$_{6}$ have been synthesized experimentally under ambient condition. So, from the chemical stability point of view we confirm that Cs$_{2}$AgFeCl$_{6}$ \cite{key_52} crystallize in cubic structures with AFM ordering while Cs$_{2}$AgCrCl$_{6}$ in hexagonal structure with ferromagnetic ordering respectively. We have simulated the chemical phase diagrams of the rest of the compounds in the series as well, and found 14 out of 27 to stabilize in a single phase (Fig. S2) \cite{key_42}. The respective competing secondary phases used for each target compound are provided in Table S4\cite{key_42}. These 14 compounds are Cs$_{2}$AgScI$_{6}$, Cs$_{2}$AgScBr$_{6}$, Cs$_{2}$AgScCl$_{6}$, Cs$_{2}$AgVBr$_{6}$, Cs$_{2}$AgVCl$_{6}$, Cs$_{2}$AgCrBr$_{6}$, Cs$_{2}$AgCrCl$_{6}$, Cs$_{2}$AgMnBr$_{6}$, Cs$_{2}$AgMnCl$_{6}$, Cs$_{2}$AgFeBr$_{6}$, Cs$_{2}$AgFeCl$_{6}$, Cs$_{2}$AgCoBr$_{6}$, Cs$_{2}$AgCoCl$_{6}$, and Cs$_{2}$AgNiCl$_{6}$. Note that, out of 14 compounds, only one iodide-based perovskite (Cs$_{2}$AgScI$_{6}$) is predicted to be stable, again confirming the difficulty in stabilizing the DPs with this anion. 
 
\begin{figure}[]
     \centering
     \includegraphics[width=1\linewidth]{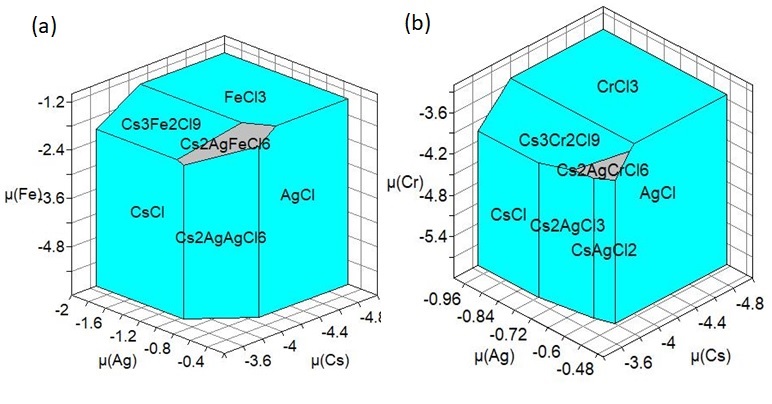}\\[-2ex]
     \caption{Chemical phase diagram of (a) Cs$_{2}$AgFeCl$_{6}$ (b) Cs$_{2}$AgCrCl$_{6}$. Grey shaded area indicates the extent of stability region for the target systems, whereas the cyan shaded area represents secondary phases. }
        \label{figure-3}
\end{figure}

\begin{table}[]%
	\centering
	\caption{\label{table-1} 
		\ Magnitude and nature of bandgap (E$_g$), difference between direct and indirect band gap ($\Delta$) and local magnetic moment on T-atoms ($m_\text{T}$) of 14 chemically stable Cs$_{2}$AgTX$_{6}$ compounds. $\uparrow$ and $\downarrow$ stands for the spin up and down channels respectively. The description within parenthesis corresponding to every system indicates their stable phase.}
	\begin{ruledtabular}
		\begin{tabular}{llccc}
			\bf{Cs$_{2}$AgTX$_{6}$}	& \bf{Band gap (E$_g$)} & \bf{$\Delta$} & \bf{$m_{\text{T}}$} \\
			& \bf{ (eV}&\bf{ (eV)} & \bf{\textbf{($\mu_B$)}}\\
			\hline
			 Cs$_{2}$AgScCl$_{6}$ (Hex NM) & 3.64 (indirect) & 0.01 & - \\  
                                   &                 &      &   \\
    Cs$_{2}$AgVCl$_{6}$ (Hex FM) & ($\uparrow$) 2.40 (indirect) & 0.01 & 1.93 \\
                                                  & ($\downarrow$) 2.46 (indirect) & 0.07 &  \\
                                  &                 &      &   \\
    Cs$_{2}$AgCrCl$_{6}$ (Hex FM) & ($\uparrow$) 1.69 (indirect) & 0.02 & 3.15 \\
                                                   & ($\downarrow$) 3.51 (indirect) & 0.07 &  \\
                                     &                 &      &   \\
    Cs$_{2}$AgMnCl$_{6}$ (Cubic FM) & ($\uparrow$) metallic & - & 4.3 \\
                                                 & ($\downarrow$) 3.81 (direct) & - &  \\    
                                    &                 &      &   \\
    Cs$_{2}$AgFeCl$_{6}$ (Cubic AFM) & 1.17 (direct) & - & 4.12 \\
                                      &                 &      &   \\
    Cs$_{2}$AgCoCl$_{6}$ (Hex NM) & 1.27 (indirect) & 0.03 & - \\
                                      &                 &      &   \\
    Cs$_{2}$AgScBr$_{6}$ (Hex NM) & 2.89 (indirect) & 0.03 & - \\
                                      &                 &      &   \\
    Cs$_{2}$AgVBr$_{6}$ (Hex FM) & ($\uparrow$) 1.95 (indirect) & 0.13 & 1.99 \\
                                                  & ($\downarrow$) 3.00 (indirect) & 0.11 &  \\
                                       &                 &      &   \\                                          
    Cs$_{2}$AgCrBr$_{6}$ (Hex FM) & ($\uparrow$) 1.09 (indirect) & 0.05 & 3.33 \\
                                                  & ($\downarrow$) 3.19 (indirect) & 0.09 &  \\
                                       &                 &      &   \\                                            
    Cs$_{2}$AgMnBr$_{6}$ (Cubic FM) & ($\uparrow$) metallic & - & 4.32 \\
                                                  & ($\downarrow$) 2.87 (direct) & - &   \\     
                                       &                 &      &   \\                                            
    Cs$_{2}$AgFeBr$_{6}$ (Cubic AFM) & 0.64 (direct) & - & 3.98 \\
                                       &                 &      &   \\    
    Cs$_{2}$AgCoBr$_{6}$ (Hex NM) & 0.96 (indirect) & 0.05 & - \\
                                      &                 &      &   \\    
    Cs$_{2}$AgNiBr$_{6}$ (Hex FM) & ($\uparrow$) metallic & - & 1.66 \\
                                                  & ($\downarrow$) 1.93 (direct) & 0.03 & \\    
                                       &                 &      &   \\                                             
    Cs$_{2}$AgScI$_{6}$ (Hex NM) & 2.51 (indirect) & 0.08 & - \\
		\end{tabular}
	\end{ruledtabular}
\end{table}

\begin{figure}[]
     \centering
     \includegraphics[width=1\linewidth]{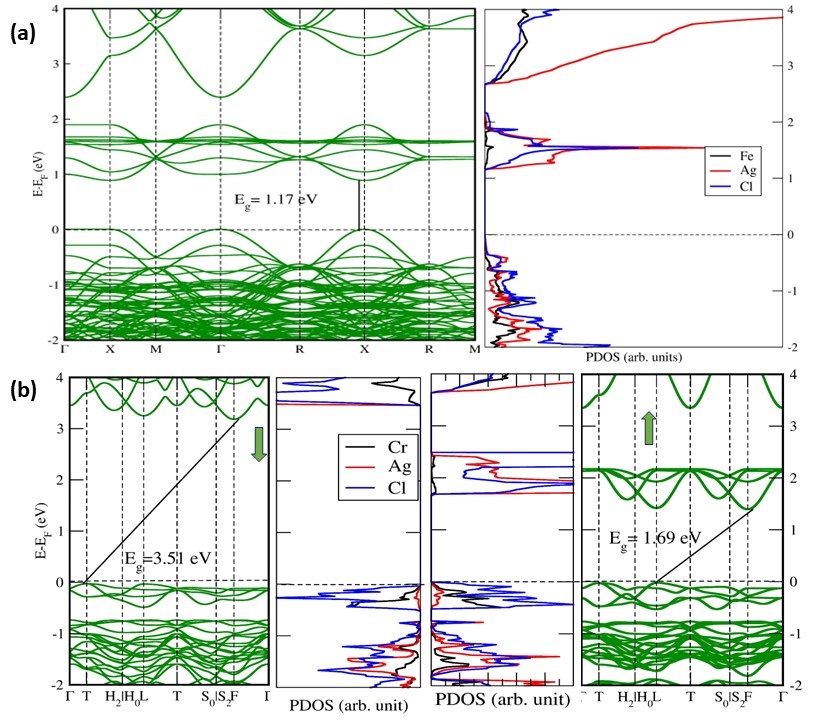}\\[-2ex]
     \caption{ Spin resolved band structure and  partial density of states (PDOS) for (a) cubic AFM Cs$_{2}$AgFeCl$_{6}$ and (b) hexagonal FM Cs$_{2}$AgCrCl$_{6}$. The former is an AFM semiconductor with a band gap of 1.17 eV, while the later is a FM semiconductor with band gaps 3.51 and 1.69 eV for spin $\downarrow$ and $\uparrow$ channels respectively } 
    \label{figure-4}
\end{figure}

\subsection{Electronic structure and magnetic properties}
Table \ref{table-1} shows the band gap (E$_g$), difference between direct and indirect band gap ($\Delta$) and atom projected magnetic moments on the transition elements (T) for all the 14 stable compounds, as mentioned in the previous section. Similar details for all the 27 compounds (Cs$_{2}$AgTX$_{6}$) are presented in {Table S5} (see supplementary) \cite{key_42}. The energetically most stable structure and the corresponding magnetic phase for each compound is mentioned within the parenthesis of the first columns of Table 1 and S5. In order to correctly capture the effect of electron-electron correlation arising out of transition elements, we have applied an onsite Hubbard U correction for each compound which is calculated self consistently. Upon Inclusion of Hubbard potential, the degeneracy of d-states found around the Fermi level is lifted in some of the examined compounds resulting in the increase of band gap. As follows from the data in Table 1, these set of compounds shows diverse electronic/magnetic properties ranging from nonmagnetic metals/semiconductors to ferromagnetic half metals to antiferromagnetic semiconductors and ferromagnetic semiconductors. 
 Cs$_{2}$AgFeX$_{6}$ (X= Cl, Br, I) show direct nature of band gap (E$_g$), with E$_g$ values ranging from 1.17 eV (Cl) to 0.64 eV (Br) to 0.12 eV (I). Though, other compounds show indirect nature of band gap, the difference ($\Delta$) between the direct (E$_g^d$) and indirect (E$_g$) gaps is quite small (ranging from 0.01-0.09 eV). As expected, the band gap value decreases owing to the decrease in the electronegativity of the halogen elements, as observed in other halide perovskites as well \cite{key_57}. The magnetism in some of these compounds mainly arises from the partially filled transition d-elements at T site{( Table 1)}. Based on these electronic/magnetic properties, one can classify these materials into different categories of compounds with (1)- moderate band gap values, (2) large band gap values, (3)-  metallic nature in one spin channel and semiconducting in the other (half-metals). Such varying properties can make these materials useful for various applications such as photovoltaics, photo(electro)catalysis and spintronics, which needs further in-depth analysis.(Section-3 of the SI for more details). 
 
To explore the potential of these materials for {the} magneto-photovoltaic application, two experimentally synthesized compounds Cs$_{2}$AgFeCl$_{6}$ \cite{key_24} and Cs$_{2}$AgCrCl$_{6}$ \cite{key_25} which show two distinct magnetic ordering (AFM and FM respectively) were chosen for deeper examination. Figures \ref{figure-4}(a) and \ref{figure-4}(b) show the bulk electronic band structures of these two {materials}. The magnetic ordering preferably shows up at low temperature which to the best of our knowledge, has not been experimentally testified. Our simulation confirms Cs$_{2}$AgFeCl$_{6}$ to be antiferromagnetic with a band gap of around 1.17 eV which agrees fairly well with experimental value of  1.55eV \cite{key_24}. Both valence band maxima (VBM) and conduction band minima (CBM) lie at a common high-symmetry X-point resulting in {a} direct band gap.  From the partial density of states (pDOS), one can see that the CBM is mostly composed of {the} Fe-d states hybridized with some of the Cl-p states with a slight contribution from {the} Ag-d states. {The} VBM is mostly composed of {the} Cl-p and Ag-d states. The strong hybridization between these states near the VBM is responsible for strong band dispersion leading to low effective mass of holes and hence higher mobility of holes as compared to electrons, indicating the p-type semiconducting behaviour of Cs$_{2}$AgFeCl$_{6}$. Under constant relaxation time approximation, we have simulated the carrier mobilities of Cs$_{2}$AgFeCl$_{6}$ and Cs$_{2}$AgCrCl$_{6}$.  Figure S8 (see supplementary) \cite{key_42} shows a comparative plot of electron and hole mobilities of Cs$_{2}$AgFeCl$_{6}$ and Cs$_{2}$AgCrCl$_{6}$ across varying temperature range at a carrier concentration of 10$^{10}$ /cm$^3$. For comparison, carrier mobilities for a well studied DP Cs$_{2}$AgBiBr$_{6}$ is also shown. The Cs-s states are  scant around the Fermi level and dominantly lie deep down the conduction band and hence play a passive role in defining the electronic properties. Due to the AFM ordering, the net magnetization of the unit cell is zero while the atom-projected Fe moment is $\sim$4.12 $\mu_B$. In the case of the Br and the  I based analogues of this material, the nature of {the} orbital contributions are similar, but the band gap decreases as expected.Next we consider Cs$_{2}$AgCrCl$_{6}$ which is a ferromagnetic semiconductor ( Figure \ref{figure-4}(b) ). It has two different semiconducting gaps ($E_g$) in two spin channels. In {the} spin down channel, the CBM lies at F high symmetry k-point while {the} VBM lies at T-point, {producing} an indirect band gap ($E_g$=3.15 eV). In {the} spin up channel, the CBM also lies at F-point but the VBM occur at L-point, with an indirect band gap of 1.69 eV. From {the} pDOS plot, one can notice a hybridization between the Cl-p and Ag-d states in valence band in both the spin channels. In {the} spin up channel the conduction band is mostly composed of the  {Cr-d} states and some contribution from {the} p-states.  In both the spin channels, {the} Ag-d states reside near the valence band because it has full d-valence electrons. The d-orbitals of Cr are half filled with three unpaired electrons giving rise to a net moment of $\sim$3.1$\mu_B$. 
The electronic band structure along with the DoS for the rest of the chemically stable compounds (Cs$_{2}$AgTX$_{6}$) are given in SI (see Figs. S4-S7) {\cite{key_42}. Robust chemical stability, optimal band gap and varying magnetic orderings makes these materials potential candidates for magneto-photovoltaics applications.

\begin{figure}[]
     \centering
     \includegraphics[width=1\linewidth]{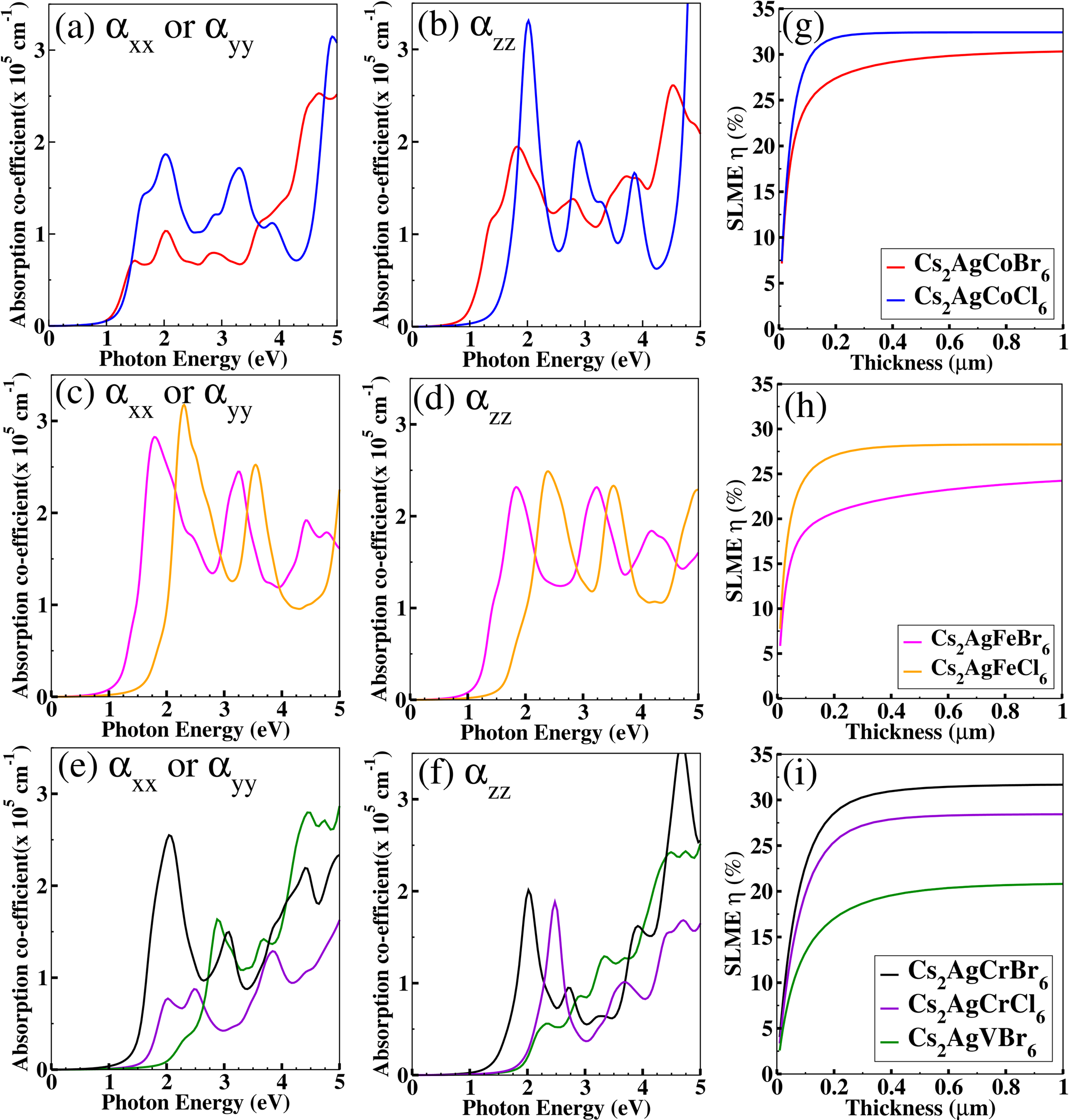}
     \caption{\textcolor{black}{Absorption coefficient($\alpha_{ii}$) along in-plane and out-of-plane direction for selected (a,b) non-magnetic, (c,d) anti-ferromagnetic, and (e,f) ferromagnetic systems. Spectroscopic limited maximum efficiency (SLME) of the same set of (g) non-magnetic, (h) anti-ferromagnetic, and (i) ferromagnetic systems.}}
        \label{figure-5}
\end{figure}
\subsection{Optical properties}
Hybrid halide perovskites have proven to be potential candidates for photovoltaic applications owing to various factors like excellent absorption coefficients and suitable band gaps. The optical absorption coefficient ($\alpha$) is calculated from the dielectric function using the following equation -
\begin{equation}
  \alpha(\omega) =\frac{\sqrt{2}\omega}{c} \sqrt{Re(\epsilon_1(\omega^2)) + \Im m(\epsilon_2(\omega^2))} - Re(\epsilon_1(\omega^2))
\end{equation}
where $c$ is the speed of light, $\epsilon_{1}$ and $\epsilon_{2}$ represent the real and imaginary parts of the dielectric function, and $\omega$ is the frequency. $\epsilon_{2}$ was calculated by using independent particle approximation \cite{key_58} and $\epsilon_{1}$ {was} calculated from $\epsilon_{2}$ using Kramers–Kronig relation \cite{key_59} In case of hexagonal structure, the dielectric tensor is diagonal with two different components: $\epsilon_{2\alpha_{zz}}$ (along the z-axis of the crystal), and $\epsilon_{2\alpha_{xx}}$ and $\epsilon_{2\alpha_{yy}}$ (along any direction in the plane perpendicular to the hexagonal z-axis).
Based on the optimal band gap values in the visible range, few of the compounds viz., Cs$_{2}$AgCoI$_{6}$, Cs$_{2}$AgCoBr$_{6}$, Cs$_{2}$AgCoCl$_{6}$, Cs$_{2}$AgFeBr$_{6}$, Cs$_{2}$AgFeCl$_{6}$, Cs$_{2}$AgCrI$_{6}$, Cs$_{2}$AgCrBr$_{6}$, Cs$_{2}$AgCrCl$_{6}$, and Cs$_{2}$AgVBr$_{6}$ {were} found to be promising for {the} PV applications.{The optical absorption of these compounds are relatively} high ($>3\times10^5 cm^{-1}$) for most of the compounds, as shown in Fig. \ref{figure-5}(a-f). The onset of these absorption curves corroborates with their respective band gaps (Table 1). For example in the case of Cs$_{2}$AgFeCl$_{6}$, one can observe that optical absorption onset is around 1.17 eV which matches our simulated electronic band gap. In order to quantify the power conversion efficiency, a parameter known as spectroscopic limited maximum efficiency (SLME) introduced by Yu et al \cite{key_43} is estimated (see Sec. 3.1 of SI for more details). For SLME, the input parameters are band gap, absorption coefficient, standard solar spectrum and the thickness. Figures \ref{figure-5}(g,h,i) show the simulated SLME for all the compounds shortlisted for PV application. The calculated SLME  at 1 $\mu$m absorber layer thickness is  23.09\%, 30.57\%, 32.41\%, 25.52\%, 28.82\%, 17.42\%, 31.72\%, 28.45\%, and 20.93\% for Cs$_{2}$AgCoI$_{6}$, Cs$_{2}$AgCoBr$_{6}$, Cs$_{2}$AgCoCl$_{6}$, Cs$_{2}$AgFeBr$_{6}$, Cs$_{2}$AgFeCl$_{6}$, Cs$_{2}$AgCrI$_{6}$, Cs$_{2}$AgCrBr$_{6}$, Cs$_{2}$AgCrCl$_{6}$, and Cs$_{2}$AgVBr$_{6}$ respectively. Interestingly, all these materials have high values of SLME even at smaller absorber layer thickness. This, along with their suitable band gap and high absorption coefficient, makes them potential candidates for thin film solar cells. The highest value of SLME is achieved for Cs$_{2}$AgCoCl$_{6}$, which could serve as the best material for PV applications.  In particular,  Cs$_{2}$AgCrBr$_{6}$, which is a ferromagnetic semiconductor, exhibits high value of absorption coefficients and SLME ($\sim$31\%), and hence can be a promising candidate for magneto-photovoltaic applications.
Yet another interesting candidate is Cs$_{2}$AgFeCl$_{6}$ which shows highly dispersive bands around the high symmetry X-point (Fig. \ref{figure-4}(a)) implying low effective mass of hole charge carriers and high hole mobility as aforementioned. Herein, both VBM and CBM are dominated by T-d states located at X point giving rise to direct band gap. In contrast, for Cs$_{2}$AgCrCl$_{6}$, two peaks are observed in the optical absorption spectrum which corresponds to two different band gaps  in the two spin channels. These two peaks can be assigned to the d-d transitions at 3.51 eV and 1.69 eV respectively. With varying halides (Cl to Br to I), there is a significant shift in the  optical absorption onset indicating a change in the electronic band gaps. 

\section{Conclusion}
\textcolor{black}{The interplay of magnetic and  optical properties can lead to a new avenue for finding candidate materials with promising photovoltaic (PV) performance. The purpose of this manuscript is to explore the said objective in a new family of lead free magnetic DPs Cs$_{2}$AgTX$_{6}$ (where T = Sc, Ti, V, Cr, Mn, Fe, Co, Ni, Cu and X = Cl, Br, I).} The combination of two transition elements (Ag and T) offers a wide range of intriguing magneto-optoelectronic properties. Out of 27 compounds, seven prospective compounds Cs$_{2}$AgCoBr$_{6}$, Cs$_{2}$AgCoCl$_{6}$, Cs$_{2}$AgFeBr$_{6}$, Cs$_{2}$AgFeCl$_{6}$, Cs$_{2}$AgCrBr$_{6}$, Cs$_{2}$AgCrCl$_{6}$, and Cs$_{2}$AgVBr$_{6}$ have shown visible-light driven band gaps, good absorption coefficients and high power conversion efficiency substantiating their potential as potential PV absorbers. Rest of the compounds in this series have the potential to  show significant promise in the field of photo(electro)catalysis and spintronics, based on their electronic properties. The driving mechanism for the distinct magnetic properties in these compounds is hybridization and super-exchange. Amongst the magnetic compounds in this series, Cs$_{2}$AgCrI$_{6}$, Cs$_{2}$AgCrBr$_{6}$, Cs$_{2}$AgCrCl$_{6}$, and Cs$_{2}$AgVBr$_{6}$ stabilize in hexagonal symmetry with ferromagnetic or antiferromagnetic ordering. This can incite broken time-reversal or inversion symmetry and lead to bulk spin photovoltaic effect, thereby enhancing the PV performance due to non-linear optical effects. For instance, large photoconductivity has been reported in CrI$_{3}$ due to the magnetism-mediated asymmetry in its antiferromagnetic ordered structure. Such asymmetry has paved a unique way to explore the novel properties of magnetic materials for anomalous photovoltaic effect (APVE). Hence, the insights on magnetic DPs Cs$_{2}$AgTX$_{6}$, reported in our work can channelise the advancement of perovskites  in the field of magnetic/spin APVE. The theoretical reaffirmation of the synthesizability of our proposed compounds paves way to also expedite experimental research in this impending class of family Cs$_{2}$AgTX$_{6}$.


\section*{Acknowledgment}
A.A acknowledges computing facility (spacetime2) provided by IIT Bombay to support this research. S.S.P acknowledges the computational support by MASSIVE HPC facility (www.massive.org.au) and the Monash eResearch Centre and eSolutions-Research Support Services through the use of the MonARCH HPC Cluster.

\section*{Author Contributions}
M.N and S.S.P contributed equally to this work.

\nocite{*}
\bibliography{pra_ref} 

\end{document}